\documentclass[reprint,prd]{revtex4-1}

\usepackage[T1]{fontenc}
\usepackage[utf8x]{inputenc}
\usepackage[brazil, english]{babel}
\usepackage{amsmath}
\usepackage{amsfonts}
\usepackage{amssymb}
\usepackage{tensor}
\usepackage[amssymb]{SIunits}
\usepackage{mathrsfs}
\usepackage{latexsym}
\usepackage{graphicx}
\usepackage{nicefrac}
\usepackage{hyperref}
\usepackage{multirow}
\usepackage{paralist}

\newcommand{\ud}{\ensuremath{\mathrm{d}}}
\def\coxa{{\Huge
C\kern-.1667em\lower.5ex\hbox{O}\-X\kern-.1667em\lower.5ex\hbox{A}\@}%
\index{CoXa}
}

\newcounter{sol}

\hypersetup{
pdftitle = {Realistic fluids as source for dynamically accreting black holes in a cosmological background},
pdfauthor = {Daniel C. Guariento, Michele Fontanini, A. M. da Silva, Elcio Abdalla},
colorlinks = true,
breaklinks = true
}

\begin{document}

\title{Realistic fluids as source for dynamically accreting black holes in a cosmological background}

\author{Daniel C. Guariento}
\email{carrasco@fma.if.usp.br}

\author{Michele Fontanini}
\email{fmichele@fma.if.usp.br}

\author{Alan M. da Silva}
\email{amsilva@fma.if.usp.br}

\author{Elcio Abdalla}
\email{eabdalla@fma.if.usp.br}

\affiliation{Instituto de Física, Universidade de São Paulo,\\
Caixa Postal 66.318, 05315-970, São Paulo, Brazil}

\begin{abstract}

We show that a single imperfect fluid can be used as a source to obtain a mass-varying black hole in an expanding universe. This approach generalizes
the well-known McVittie spacetime, by allowing the mass to vary thanks to a novel mechanism based on the presence of a temperature gradient. This fully
dynamical solution, which does not require phantom fields or fine-tuning, is a step forward in a new direction in the study of systems whose local
gravitational attraction is coupled to the expansion of the universe. We present a simple but instructive example for the mass function and briefly
discuss the structure of the apparent horizons and the past singularity.

\end{abstract}


\maketitle


The McVittie metric is a controversial solution to Einstein's equations \cite{mcvittie-1933} modelling a black hole in a cosmological Friedmann-Lema\^{\i}tre-Robertson-Walker (FLRW) background,
carrying now almost 80 years of debates about its physical interpretation. It is a particular case, and the first to be discovered, of a class of
metrics which describe a spherically symmetric expanding spacetime filled by a comoving shear-free fluid
\cite{stephani-exact,*kustaanheimo-1947,bolejko-2011}. Several papers \cite{kaloper-2010,carrera-2010,lake-2011} pointed out that the central object
in the McVittie solution has the correct prerequisites to describe a black hole. Despite this, a non uniform agreement on the topic remains in the
literature \cite{nolan-1998,*nolan-1999,*nolan-1999a,gao-2008,*faraoni-2009}.

Solutions of this kind are of interest for those who study processes both at compact object and cosmological scales, for they bridge between the two
realms, giving a deeper understanding of cross-scale problems. Probably for this very reason many attempts in generalizing the McVittie solution have
been made during the years, as can be seen in \cite{faraoni-2012} and references therein. Of course, one of the features searched for in
generalizations of this solution is the possibility of accreting mass on the compact object
\cite{ademir-plb-2010,*guariento-2011,*babichev-2005,*guariento-2007}. Despite being a quite reasonable request, especially when one thinks of
physical objects such as stars and black holes, the introduction of accretion has been shown to lead to extremely challenging difficulties
\cite{babichev-2012,*dokuchaev-2011}, even in simpler cases as in Vaidya metric and perturbations thereof
\cite{shao-2004,*abdalla-2006,*abdalla-2007}.

In this Letter, by completely abandoning the assumption of a perfect fluid, an approach that has shown to lead to interesting results in many fields
\cite{faraoni-2007,abreu-2007,herrera-1997,*herrera-2004,*herrera-2008}, but which has not received enough attention, we
find that one can in fact describe accretion (or evaporation) of a black hole in an FLRW background. Although perfect fluids are a valuable tool to
study the universe at very large scales, they are not enough in general to match observations, as is shown for instance in
\cite{nagai-2006,navarro-1996} where radiative cooling is required in numerical simulations to agree with data on density profiles in clusters. The
importance of imperfect fluids has been noticed even at cosmological scales, where they have been proven useful in different contexts
\cite{sussman-1999,*sussman-2008,brevik-2005,*brevik-2006}.

We do not focus on proving that a generalized McVittie metric actually describes a central black hole in an FLRW universe, given that the analysis in
\cite{kaloper-2010} and \cite{lake-2011} which guarantees this fact is still applicable. Rather, we show how such a solution can be constructed,
analyze some of its properties and present what could be called a toy model for the black hole evolution. We stick to a simple example to keep
computational difficulties at bay, while we recognize that the imperfect fluid formulation we are presenting can accommodate for a vast plethora of
behaviors which we plan to analyze in future works. In short, we look at a very slowly changing mass between two constant values in an asymptotically
de~Sitter cosmology.


The McVittie metric \cite{mcvittie-1933} is defined by the line element
\begin{equation}\label{gmv}
  \ud s^2 = -\frac{\left(1 - \frac{m}{2 a r} \right)^2}{\left(1 + \frac{m}{2 a r}\right)^2} \ud t^2 + a^2  \left(1 + \frac{m}{2 a r}\right)^4 \left(
    \ud r^2 + r^2\ud \Omega^2 \right) \,,
\end{equation}
where $a = a(t)$ and $m$ is a constant. It is a solution to Einstein's equations initially proposed to describe a central object in a
cosmological background driven by a perfect fluid with homogeneous density and inhomogeneous pressure. In the limit in which $H \equiv
\nicefrac{\dot{a}}{a}$ goes to a constant, it reduces to the Schwarzschild--de~Sitter solution \cite{kottler-1918}, and in the limit $\nicefrac{m}{2 a
  r} \ll 1$ it reduces to a perturbed FLRW universe with zero curvature. The $m$ parameter represents the contribution to the Misner--Sharp mass
coming from the central inhomogeneity.

Among the main characteristics of the spacetime described by this metric is the presence of a spacelike inhomogeneous big bang singularity at $m = 2 a
r$, which lies in the causal past for $\dot{a} > 0$. The metric has a null or spacelike FLRW future infinity at large $r$ and $t$ and two apparent
horizons which are anti-trapping surfaces at finite times. The one at the lower value of $r$ is well behaved at late times in the presence of a positive cosmological
constant and, for $t = \infty$, becomes the event horizon of a Schwarzschild--de~Sitter black hole, which is reachable in a finite proper time
\footnote{The fate of this surface in a different limit for $H$ is still a source of debate \cite{kaloper-2010,lake-2011,bolejko-2011}.}. Therefore,
one can say that in the case in which the scale factor $a$ asymptotes de~Sitter, the metric \eqref{gmv} describes a black hole embedded in an FLRW
spacetime \cite{kaloper-2010}.


To generalize the McVittie metric, we consider a time-varying mass for the central object, namely $m = m(t)$  in \eqref{gmv}
\cite{faraoni-2007,anderson-2011}. Several difficulties are introduced by this apparently small change; the first immediate one is to find a
reasonable fluid to enter in the energy-momentum tensor.


The nonzero components of the Einstein tensor for the generalized McVittie line element acquire extra contributions with respect to the original case
depending on $\dot{m}$. They read
\begin{subequations}\label{estn}
\begin{gather}
  \label{estn00}
  G\indices{^t_t} = -3 \left[ \frac{\dot{a}}{a} + \frac{2 \dot{m}}{2 a r - m} \right]^2 \,, \displaybreak[3]\\
\label{estn01}
  G\indices{^t_r} = -8 a \dot{m} \frac{2 a r + m}{\left(2 a r - m \right)^3} \,,\\
\label{estn11}
  G\indices{^r_r} = G\indices{^\theta_\theta} = G\indices{^\phi_\phi} = G\indices{^t_t} - 2 \frac{2 a r + m}{2 a r -m} \frac{\ud}{ \ud
    t} \sqrt{-\frac{ G\indices{^t_t}}{3}} \,.
\end{gather}
\end{subequations}

The off-diagonal term, together with the fact that $G\indices{^r_r}$ and $G\indices{^\theta_\theta}$ have to be equal, puts a stringent constraint on
the choice of fluid. If a single perfect fluid is used as a source for this metric, the equality of the diagonal terms implies that the fluid has to
be comoving \cite{guariento-tese-2010}. This in
turn implies that the off-diagonal term is zero, and therefore that $\dot{m}$ has to vanish. It follows then that no single perfect fluid
description can be used as a source for the generalized McVittie.

The problem can be alleviated with the addition of a second perfect fluid, which is forced though to have a phantom equation of state ($p
< -\rho$), and thus carry all the problems associated with this kind of field (see, for instance, \cite{caldwell-2002,bamba-2012} for an
introduction on the topic). Moreover the two fluids are required to be connected by a quite unnatural balancing equation.

Therefore, to find a suitable single-fluid interpretation for the metric \eqref{gmv}, we require more complexity and introduce heat transport and
viscosity. Considering a comoving fluid, which, together with the spherical symmetry of the metric, implies the vanishing of the shear viscosity, the
most general form for an imperfect fluid which is compatible with thermodynamical requirements up to first order in the gradients is
\cite{weinberg}
\begin{multline}\label{temini-geral}
  T \indices{^{\mu\nu}} = \left( \rho + p \right) u^{\mu} u^{\nu} + p \ g \indices{_{\mu\nu}} - \zeta H \indices{^{\mu\nu}}
  u\indices{^{\gamma}_{;\gamma}} \\
- \chi \left( H \indices{^{\mu\gamma}} u^{\nu} + H \indices{^{\nu\gamma}} u^{\mu} \right) Q_{\gamma} \,,
\end{multline}
with
\begin{equation}
H \indices{_{\mu\nu}} = g \indices{_{\mu\nu}} + u_{\mu} u_{\nu} \,,\;
Q_{\mu} = \partial_{\mu} T + T u\indices{_{\mu ;\gamma}} u^{\gamma} \,,
\end{equation}
and where $u^\mu$ is the four-velocity, $T = T(x^\mu)$ the fluid temperature, $\chi$ the heat conductivity and $\zeta$ the bulk viscosity.

The energy-momentum tensor thus takes the form
\begin{subequations}
\begin{gather}\label{T00}
  T\indices{^t_t} = -\rho \,,\\
  \label{T01}
  T\indices{^t_r} = -\chi \left[ \left( \frac{2 a r + m}{2 a r -m}
    \right) \partial_r T + \frac{4 m a T}{(2 a r - m)^2} \right] \,,\\
  \label{Tii}
  T\indices{^r_r} = T\indices{^\theta_\theta} = T\indices{^\phi_\phi} = p - 3 \zeta \left( \frac{\dot{a}}{a} + \frac{2 \dot{m}}{2 a r - m} \right)
  \,.
\end{gather}
\end{subequations}


The off-diagonal component of Einstein's equations given by \eqref{estn01} and \eqref{T01} can be solved to find the radial dependence of the
temperature, obtaining a function that relates the evolution of the temperature profile to the mass,
\begin{equation}
  T = \frac{1}{\sqrt{-g\indices{_{tt}}}} \left[ T_{\infty} (t) + \frac{\dot{m}}{4 \pi \chi m} \ln \left( \sqrt{-g\indices{_{tt}}} \right)\right] \,,
\end{equation}
where $T_{\infty} (t)$ is an arbitrary function of time that represents the value of the temperature at spatial infinity. It is interesting to note
that if $\dot{m} = 0$ the fluid temperature is just $\frac{T_{\infty} (t)}{\sqrt{-g\indices{_{tt}}}}$, which is equivalent to saying that it is in
thermal equilibrium \cite{zheng-1997}. Therefore, the time-dependent black hole mass may be seen as a direct consequence of the fluid being out of
equilibrium. It is thanks to this energy transfer mechanism involving heat flow that the fluid can still be comoving and admit nonzero off-diagonal
components  in the metric, therefore satisfying what Carrera and Giulini call the spatial Ricci-isotropy of the Einstein tensor \cite{carrera-2010},
expressed in the first part of equation \eqref{estn11}.

Using the remaining Einstein's equations, we can extract solutions for the fluid energy density in terms of the black hole mass, which can always be
written as
\begin{equation}\label{rho}
\rho(r,t) = \frac{3}{8 \pi} \left[\frac{2 \dot{m}}{\left(2 a r -m \right)} + H \right]^2 \,.\\
\end{equation}
Of course now the freedom in our choice is expressed by the almost arbitrary function $\dot{m}$. The requirements for it to be a consistent solution
of Einstein's equations are \begin{inparaenum}[\bfseries (i)] \item that $\dot{m}$ does not change its sign throughout the evolution, and \item if
  $\dot{a}$ and $\dot{m}$ have opposite signs, $2 \dot{m} - H (m - 2 a r)$ must have the same sign as $\dot{a}$\end{inparaenum}.

This defines a family of solutions, each of which is fully determined once $T_{\infty}$, $\dot{m}$ and $a$ are chosen. Note that these are functions of
time only, and that consequently the radial profiles are fully determined. While $T_\infty$ and $a$ describe characteristics of the fluid and of the
metric at spatial infinity, $\dot{m}$ determines the behavior of the solution at small $r$, namely the evolution of the black hole and of the
inhomogeneous part of the fluid's energy density.

The fluid pressure is fully determined if the above functions are given, and its expression  may be obtained by
substituting the solution for $m(t)$, found formally integrating the chosen $\dot{m}$, back into \eqref{estn11} and \eqref{Tii}. The pressure differs
from the original McVittie case \cite{kaloper-2010} by the terms from \eqref{estn11} that depend on $\dot{m}$ and by the bulk viscosity term from
\eqref{Tii}, which introduces a further deviation from the perfect fluid case. This last effect relates to viscous cosmology models
\cite{brevik-2005,*brevik-2006}.



Similarly to the original McVittie case, it is convenient to work in a different set of coordinates in which the radial coordinate coincides with the
``areal radius''. We define then $\hat{r}$ by
\begin{equation}
\hat{r}(t,r)=\left(1+ \frac{m}{2ar}\right)^2 a r \,,
\end{equation}
which, as in the original McVittie case, defines two branches \cite{kaloper-2010}. We choose the branch mapping $\hat{r}$ from $m = 2 a r$ at $\hat{r}
= 2 m$ to $r \to \infty$ at $\hat{r} \to \infty$. The other branch terminates on spacelike singularities both in the past and in the future
\cite{kaloper-2010} and is thus not relevant for our analysis. The line element may then be cast as
\begin{equation}\label{metricaEF}
\ud s^2 \!=\! - R^2 \ud t^2 \! + \! \left[ \frac{\ud \hat{r}}{R} \!-\! \left( \! H \! - \! M \! + \! \frac{M}{R} \right) \hat{r} \ud t \right]^2 \!\! + \! \hat{r}^2 \ud \Omega^2 ,
\end{equation}
where we have introduced the simplifying notation $R\equiv \sqrt{1-\frac{2m}{\hat{r}}}$ and $M\equiv\frac{\dot{m}}{m}$.

In order to determine the apparent horizons we compute the extrema of the area swept by a congruence of light curves. Due to spherical symmetry we only
need to focus on radial null geodesics, which satisfy $\ud s^2 = 0$. From \eqref{metricaEF} it immediately follows that for such curves
\begin{equation}\label{app-hor2}
\left( \frac{\ud \hat{r}}{\ud t} \right)_{\pm} = R \left( \hat{r} H \pm R \right) + \hat{r} M \left( 1 - R \right) \,.
\end{equation}

Since the area of the wavefront is given by $A(\hat{r},t) =4\pi \hat{r}^2$, the extrema of $A$ correspond to the solutions of $\frac{\ud\hat{r}}{\ud
  t}=0$. In principle, the full set of solutions that define the surfaces we are searching for is given by $\left(\frac{\ud \hat{r}}{\ud t}
\right)_+\left(\frac{\ud \hat{r}}{\ud t} \right)_- = 0$, which corresponds to $g_{tt} = 0$ in \eqref{metricaEF}. Once rationalized, this is an eighth order
equation, as opposed to the sixth order equation one encounters in the original McVittie case.

In the branch we are using $0< R < 1$, and the problem simplifies considerably when an accreting  black hole ($M > 0$) in an expanding universe  ($H >
0$) is considered. In this case in fact, the outgoing null rays corresponding to the plus sign in \eqref{app-hor2} do not admit real solutions,
therefore we only consider the minus sign.


A full analysis of the causal structure is under consideration \cite{us-in-preparation-2012}. In what follows, we focus on a simple toy model where we
take the mass of the black hole to be constant both at early and at late times, and we smoothly interpolate between the two constant mass regimes at
intermediate times. This choice is dictated mostly by simplicity; a more physical choice for the mass function would require some deeper understanding
of the still not well understood accretion mechanism of self-gravitating fluids. We select our model choosing a form for the scale factor and for the
mass,
\begin{gather}
\label{hmath0}
H(t)=\frac{2}{3 t} + H_0 \,,\\
\label{msin}
m(t)=\begin{cases}
1 & t \le t_0 \,;\\
\frac{1}{2} \left[ 3 + \sin \left(\omega t + \phi \right) \right] & t_0 < t < t_1 \,;\\
2 & t \ge t_1 \,,
\end{cases}
\end{gather}
where $\omega$ and $\phi$ are appropriate parameters to have the sine monotonically connecting the two constant mass values. The function \eqref{msin}
goes smoothly (with zero derivative) from the initial to the final mass as the sine takes values from $-1$ to $+1$ in a half period. When $\dot{m}$
ceases to be zero the energy density and temperature acquire gradients toward the singularity where they themselves go to infinity. The presence of
this density gradient in the dynamical case avoids the rather artificial setup of the original McVittie, whose requirement of a homogeneous density
supported by pressure gradients was physically difficult to justify. Conversely, the pressure, besides showing discontinuities which are a feature of
the oversimplification introduced in this special case, behaves much like in the static-mass case, going to infinity at the singularity.

\begin{figure}[!htp]
  \centering
  \includegraphics[width=.45\textwidth]{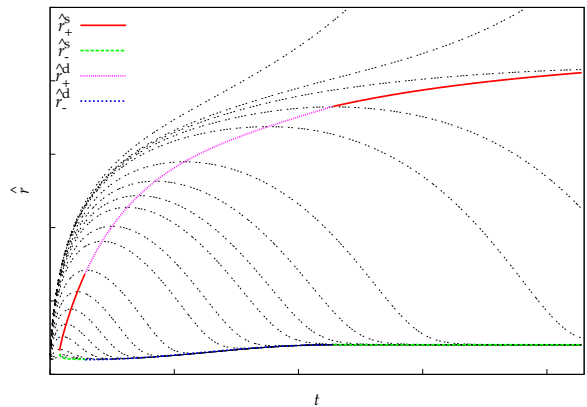}
  \caption{Apparent horizons for the dynamical McVittie metric with $H(t)$ and $m(t)$ defined in equations \eqref{hmath0} and \eqref{msin}, along with
    trajectories of radial ingoing null geodesics. $\hat{r}_+$ is the outer horizon and $\hat{r}_-$ is the inner horizon. The label ``s'' refers to
    the portions in which the generalized McVittie metric corresponds to a static-mass case with the value of the mass function given by either the
    initial or the final value in \eqref{msin}, whereas ``d'' corresponds to the new behavior only present due to the metric's dynamical evolution.}
  \label{hor-msin}
\end{figure}

In principle, the horizon equation \eqref{app-hor2} would turn out to be a fourth order equation once the square roots are eliminated. It is important
to note, though, that in the process of squaring spurious solutions can be introduced. In particular, after some work on \eqref{app-hor2} one has
\begin{equation}
\sqrt{1 - \frac{2 m}{\hat{r}}} = \frac{1 - \frac{2m}{\hat{r}} - \hat{r} M}{\hat{r} (H - M)} \,,
\end{equation}
which enforces a condition for the right-hand side to be positive. This constraint actually eliminates one of the real positive roots of the fourth order equation leaving
only two roots that we call $\hat{r}_+$ and $\hat{r}_-$. The solutions, as well as the plotted trajectories of
ingoing null geodesics obtained by numerically solving equation \eqref{app-hor2} (following  \cite{lake-2011}), are plotted in
Figure \ref{hor-msin}.

\begin{figure}[!htp]
  \centering
  \includegraphics[width=.45\textwidth]{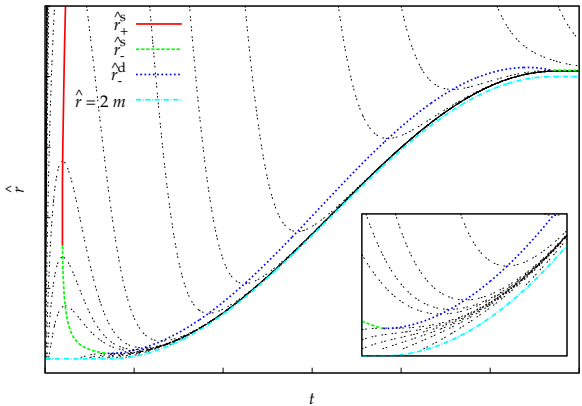}
  \caption{Lightlike trajectories of ingoing geodesics in the vicinity of the inner horizon. Although the geodesics do cross the inner apparent
    horizon as it moves, they never reach the surface $\hat{r} = 2 m$ and are eventually repelled back out of the horizon.}
  \label{hor-msin-in}
\end{figure}

Note that, for an accreting black hole, the surface $\hat{r}_-$ is traversable. This does not change the fact that it is an anti-trapping surface,
rather it means that new anti-trapping regions are appearing above it as the mass increases and the horizon moves outwards. Figure \ref{hor-msin-in}
shows this inner region in more detail.

In the original McVittie metric with $\dot{m} = 0$ the spacelike singularity at $\hat{r} = 2 m \equiv \hat{r}_*$ lies to the past of all timelike
curves if $\dot{a} > 0$ (a Big Bang) or to the future if $\dot{a} < 0$ (a Big Crunch) \cite{kaloper-2010}. This is seen by looking at the sign of
$\frac{\ud   t}{\ud \hat{r}}$ in the limit $\hat{r} \to 2 m$. In our dynamic mass case, applying such a limit to equation \eqref{app-hor2}, we find
\begin{equation}
  \lim_{\hat{r}\to 2 m}\frac{\ud t}{\ud \hat{r}} = \frac{1}{2\dot{m}} \,,
\end{equation}
with the next leading-order terms  proportional to $H$ vanishing. This is exactly the slope of the curve $\hat{r}_*=2m$, meaning that light cones
become tangent to this surface, just as in the static-mass McVittie case. The increment to the variation of $\hat{r}$ with respect to the position of
the singularity can be written as
\begin{equation}
\frac{\ud}{\ud t} \left(\hat{r} -\hat{r}_* \right) = R \hat{r} \left( H - M \right) + R^2 \left( M \pm 1 \right) \,,
\end{equation}
which for small $R$ depends on the sign of $H-M$, and where the plus and minus signs refer to outgoing and ingoing. If it is positive, as is the case in our example, every null curve will move away from
$\hat{r}_*$ as $t$ increases.


As a conclusion, we have shown that a single imperfect fluid can be used as a source to obtain the generalized McVittie metric as an exact solution to
Einstein's equations, and that the mass variation can be interpreted as a consequence of heat flow in the radial direction within the fluid. We have
worked out a simple example of an accreting black hole to reveal its main characteristics and its differences with respect to the static-mass case,
while still keeping the necessary conditions for the McVittie metric to be interpreted as a black hole at future infinity. In the
case of a slow accretion rate, the main characteristics of the McVittie metric are still present, despite the shifting position of the apparent
horizons and of the past singularity.

In the latter part of our analysis, we have actually restricted ourselves to the simplified case of slow accretion when compared to the rate of
expansion. In fact, if one moves away from this limit, we can immediately see that new features of the spacetime may emerge. For example, if one
crosses the limit $H = 2M$, some coefficients of the rationalized equation \eqref{app-hor2} will vanish, drastically changing the behavior of the
apparent horizons. Furthermore, if one crosses $H = M$ the spacelike character of the singularity is no longer guaranteed. We will address these
implications in a future work \cite{us-in-preparation-2012}.

\begin{acknowledgments}

We thank M. Lima, L. R. W. Abramo, M. Le Delliou, J. E. Horvath, E. G. M. Ferreira, C. E. Pellicer, G. I. Depetri and J. Oliveira for helpful
discussions. This work is supported by FAPESP and CNPq, Brazil. 

\end{acknowledgments}

\bibliography{referencias}

\end{document}